\documentclass[preprint,showpacs,preprintnumbers,amsmath,amssymb]{revtex4}

\usepackage{graphicx}
\usepackage{dcolumn}
\usepackage{bm}

\newcommand{\be}{\begin{equation}}
\newcommand{\ee}{\end{equation}}
\newcommand{\bea}{\begin{eqnarray}}
\newcommand{\eea}{\end{eqnarray}}

\begin{document}
\topmargin    0.2in

\preprint{}

\title{Neutron Capture Rates near $A=130$ which Effect a Global Change to the $r$-Process Abundance Distribution}

\author{R. Surman}
\affiliation{Department of Physics, Union College, Schenectady, NY 12308}

\author{J. Beun}
\affiliation{Department of Physics, North Carolina State University, Raleigh, NC 27695-8202}

\author{G. C. McLaughlin}
\affiliation{Department of Physics, North Carolina State University, Raleigh, NC 27695-8202}

\author{W. R. Hix}
\affiliation{Physics Division, Oak Ridge National Laboratory, Oak Ridge, TN 37831-6374}

\date{\today}

\begin{abstract}

We investigate the impact of neutron capture rates near the $A=130$ peak on the $r$-process abundance pattern.  We show that
these capture rates can alter the abundances of individual nuclear species, not only in the region of $A=130$ peak, but also
throughout the abundance pattern.  We discuss the nonequilibrium processes that produce these abundance changes and determine
which capture rates have the most significant impact.

\end{abstract}


\maketitle

The origin of approximately half of the solar abundances of elements heavier than the Fe group ($Z \gtrsim 26$) is the
astrophysical rapid neutron-capture process known as the $r$-process \cite{Bur57,Cam57}. While the basic
recipe for generating $r$-process elements is understood, the production of heavy nuclei through repetitions of
neutron-capture and $\beta$-decay, much remains to be learned about the nuclear physics and the astrophysics of this process.

On the astrophysics side, a number of environments are under consideration as candidate sites.  These include core collapse
supernova, e.g. \cite{Mey92,Woo94,Tak94,Beu08a}, neutron star mergers, e.g. \cite{Fre99,Mey89}, black-hole neutron star mergers,
e.g. \cite{Sur08}, gamma ray bursts, e.g. \cite{Sur06}, and shocked surface layers of O-Ne-Mg cores, e.g. \cite{Nin07,Wan03}. 
On the nuclear physics side, there is a lack of experimental data for most of the $\sim$ 3000 nuclei that participate in the
$r$-process since the majority of these nuclei are highly unstable.  With the development of radioactive beam technology, an
increasing number of these nuclei are now accessible to experiment. 

It is well known that there are two pieces of nuclear data that are important for the $r$-process: beta decay rates and nuclear masses,
for a review see \cite{Arn07}.  The former has the largest leverage on the abundance pattern since the beta decay rates
determine the time it takes to progress from the seed nuclei to the large $A$ nuclei. Thus in traditional environments, such as the
neutrino driven wind, the beta decay rates can determine whether an $A=195$ peak is formed or not.  The mass model is generally
accepted to be important as well, since the equilibrium abundances along isotopic chains are determined by neutron separation energies,
which are in turn determined by the nuclear masses.  Thus, the mass model has considerable leverage on the shape of the distribution. 

In contrast, neutron capture rates have received relatively little attention.  One might be tempted to argue that since the
actual values of the cross sections only become important during the last stages of the $r$-process, i.e. during freeze-out,
their impact is limited. However, Surman and Engel \cite{Sur01} pointed out that neutron-capture rates in the $A=195$ region can
influence the $r$-process abundances in this same region at late times.  Additionally, a number of groups
\cite{Rau05,Far06,Gor98,Gor97,Rau04} have examined simultaneous modifications of all neutron capture rates; Rauscher
\cite{Rau05} showed that such modifications can alter the time until the onset of freeze-out.  Beun et al. \cite{Beu08b} noted
that a single neutron capture rate, that of $^{130}$Sn, can affect changes throughout the $r$-process abundance pattern.  In
this paper we present calculations of the influence of neutron capture rates near the $A=130$ region on the $r$-process
abundance pattern. We find that the effect is non-negligible, affects nuclei across the entire $r$-process region, and can be
comparable to the distribution uncertainties which stem from different nuclear mass models.

Since there is little in the way of measurements for the relevant neutron capture rates, uncertainties are hard to
quantify.  A comparison of three different sets of theoretically calculated neutron capture rates \cite{Beu08b}
shows that most rates vary within a factor of 100, although a handful vary up to a factor of 1000 or more.

\begin{figure}
\includegraphics[width=16cm]{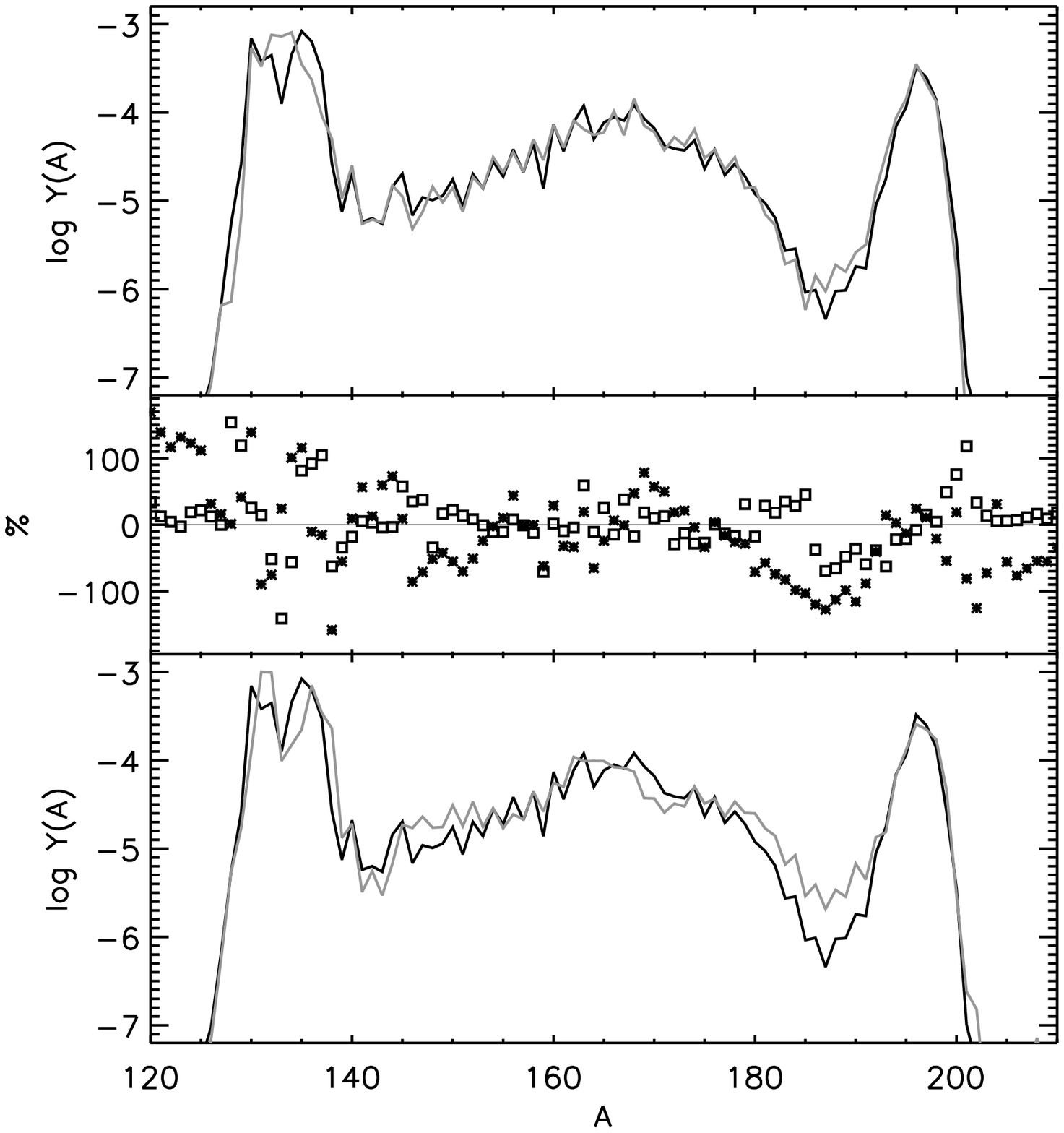}
\caption{The bottom panel shows two different
abundance patterns that were obtained using the same astrophysical conditions but
different mass models.  The top panel shows abundance patterns that were obtained
using the same mass model, but by increasing fifty neutron capture rates in the $A=130$
peak region by a factor of 100.  The middle panel shows the percent difference in abundance
for each point on the curves on the bottom panel (stars) and the top panel (squares). 
\label{fig:abund}}
\end{figure}

The bottom panel of Figure \ref{fig:abund} shows two different abundance patterns that we obtain by changing the mass model from ETFSI
\cite{Pea96} to FRDM \cite{Mol95}.  We use capture rates of \cite{Gor00} with the ETFSI masses and of
\cite{Rau00} with the FRDM masses; in both cases we use beta decay rates from M\"oller et al. \cite{Mol03}. 
We use a parameterized neutrino driven wind as in \cite{Beu08b}, with a timescale $\tau = 0.1$ s, a constant entropy per baryon of
$s/k=100$, and initial electron fraction of $Y_e = 0.26$.  In the top panel of Fig. \ref{fig:abund} we show abundance patterns using
the same astrophysical conditions, beta decay rates, and the ETFSI mass model, but with the neutron capture rates of a range of nuclei
with $128<A<138$ changed by a factor of 100.  Since abundance plots are usually made on a log scale, differences in nuclear abundances
are better seen on a linear plot.  In the middle panel we show percent differences for both scenarios.  The average change in abundance
is 43\% for the neutron capture rates and 52\% for the mass model, where the latter effect is due primarily to the differing neutron
separation energies.

\begin{figure}
\includegraphics[width=11cm]{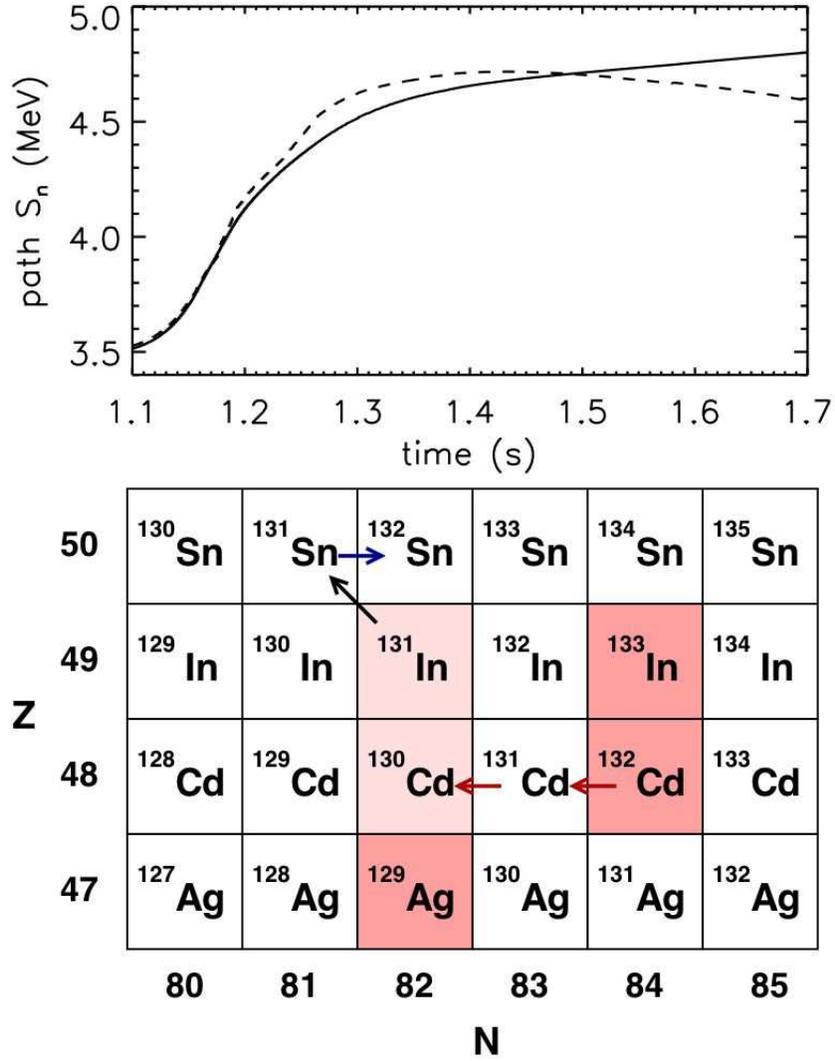}
\caption{Top panel: Shows the equilibrium (dashed line) and actual (solid line) separation energies along the
$r$-process path as a function of time for the onset of freezeout in the baseline simulation
with FRDM masses and capture rates and conditions $\tau=0.1$ s,
$Y_e = 0.26$, and $s/k = 100$.  Bottom panel: Depicts
the nuclear flow in the region of $^{131}{\rm Sn}$ and in the region of $^{131}{\rm Cd}$ for the baseline simulation. 
Shaded squares show the relevant location of the path: at the beginning of freeze-out from equilbrium $^{129}{\rm
Ag}$, $^{132}{\rm Cd}$, and $^{133}{\rm In}$ are highly populated and toward the end $^{130}{\rm Cd}$ and $^{131}{\rm
In}$ are highly populated. In color version, red arrows indicate photodissociation, blue arrows indicate neutron
capture, and the black arrow shows beta decay. 
\label{fig:schematic}}
\end{figure}

Neutron capture rates cannot influence the abundance distribution throughout most of the $r$-process when the temperature, $T$, and
neutron number density, $n_n$, are high enough for (n,$\gamma$) $\leftrightharpoons$ ($\gamma$,n) equilibrium.  This occurs when the
neutron capture rate, $\lambda_n$, and the photo-dissociation rate, $\lambda_\gamma$, between two adjacent nuclei are balanced and the
abundance distribution along an isotopic chain is determined entirely by the Saha equation.  The series of most abundant isotopes for each
element as determined by the Saha equation is called the equilibrium $r$-process path.  We determine the actual abundances and path using
a dynamical nuclear network calculation \cite{Sur01,Beu08b}.  In the dynamical calculation, each set of (n,$\gamma$)
$\leftrightharpoons$ ($\gamma$,n) rates falls out of equilibrium at different points in the simulation.  A comparison between the
equilibrium path and the actual path provides a measure of how strongly equilibrium is maintained throughout, which we quantify as
follows. We calculate an abundance averaged neutron separation energy for the system by two different methods: a) taking the abundance
weighted average of the separation energy, $S_{n,actual}$, and b) taking the equilibrium abundance weighted average of the separation
energy, $S_{n,eq}$. When the neutron capture and photo-dissociation rates are no longer sufficiently rapid to respond to changes in the
equilibrium $r$-process path resulting from changes in the temperature and/or neutron density, equilibrium fails, individual rates become
important, and $S_{n,actual}$ departs from $S_{n,eq}$, as shown in the top panel of Fig. \ref{fig:schematic}.

\begin{figure}
\includegraphics[width=16cm]{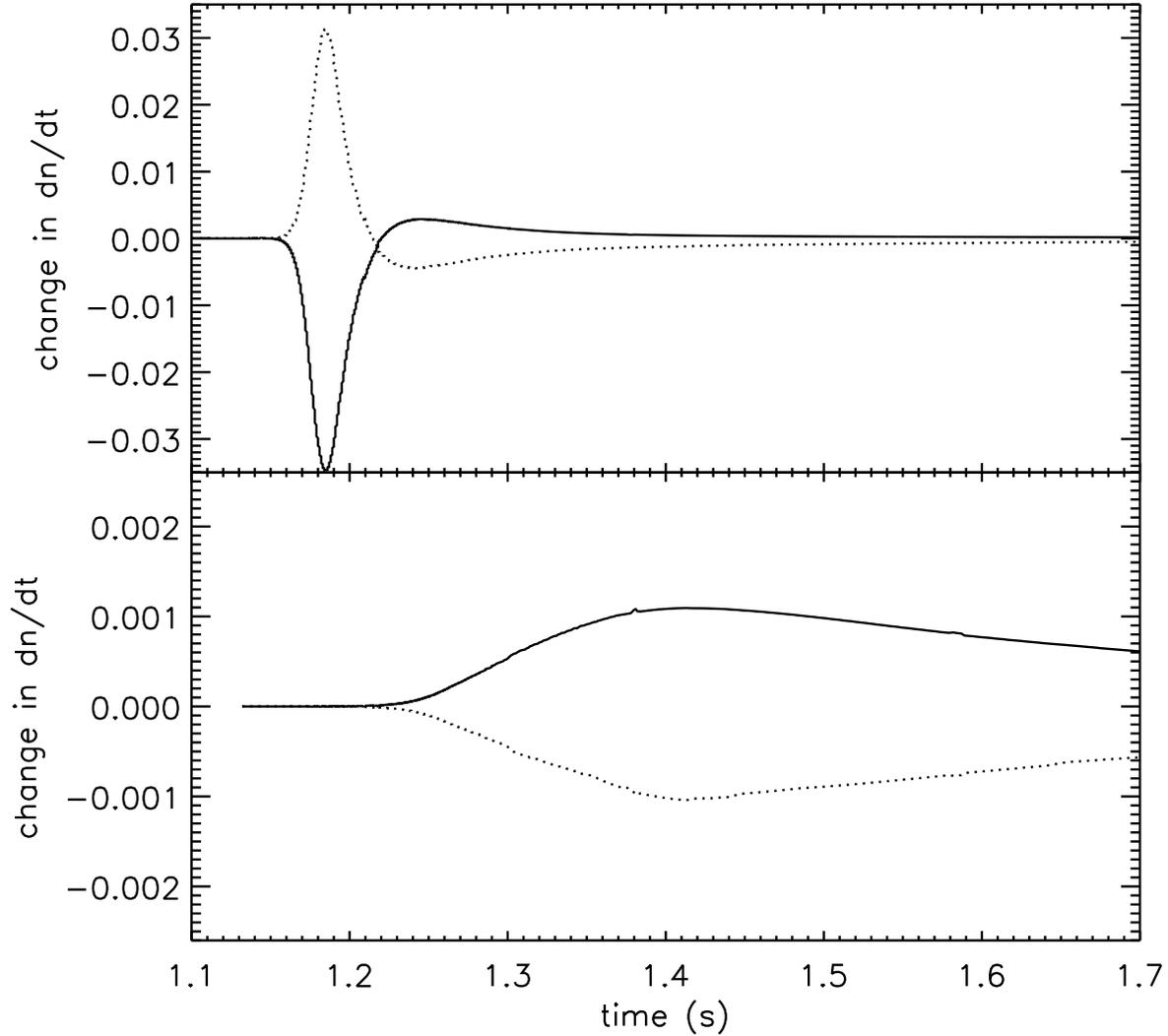}
\caption{Top panel: Shows the effect of increasing the rate of $^{131}{\rm Cd}$ by a factor of 10 relative to the baseline
simulation. The black line shows the rate of neutrons captured in the $A=130$ region relative to the baseline simulation and the
dotted line shows the rate of neutrons captured above the $A=130$ region. Bottom panel: Shows the effect of increasing the
neutron capture rate of $^{131}{\rm Sn}$ by a factor of 100 relative to the baseline simulation.  More neutrons are captured on
$^{131}{\rm Sn}$ (and therefore in the $A=130$ peak - black line), leaving less neutrons to be captured in the region above
the $A=130$ peak (dotted line). 
\label{fig:capture}}
\end{figure}

On the left hand side of this panel, the separation energies are rising as the system moves toward stability at late times.  However
the photo-dissociation rates are not rapid enough to maintain equilibrium so $S_{n,actual}$ cannot keep pace with $S_{n,eq}$.  At this
point individual photo-dissociation rates can become important. In particular, the photo-dissociation of nuclei that are highly
populated can influence the overall neutron number density.  As an example, we consider $^{132}$Cd in the baseline simulation of Fig.
\ref{fig:schematic}.  $^{132}$Cd is along the equilibrium and actual $r$-process path while equilibrium obtains; as $S_{n,eq}$ rises as
described above, the equilibrium path shifts to $^{130}$Cd.  However, the photo-dissociation rate of $^{132}$Cd is too slow to fully
shift the actual path in the baseline simulation.  The photo-dissociation rate of a nucleus $(Z,A+1)$ depends on the capture rate of
the nucleus $(Z,A)$:
\begin{equation} 
\lambda_\gamma \propto T^{3/2} \exp\left(-{\frac{S_n}{kT}}\right) \langle \sigma v \rangle_{Z,A}
\end{equation} 
where $\langle \sigma v \rangle_{Z,A}$ is the averaged value of the neutron capture cross section, $\sigma$, and the
relative velocity, $v$.  We therefore perform a simulation where the capture rate of $^{131}$Cd is increased by a
factor of 10 and compare with the baseline simulation. With the increased rate, at late times $^{132}$Cd is able to
photo-dissociate to $^{131}$Cd, which in turn promptly photo-dissociates to $^{130}$Cd, as depicted in the bottom panel
of Figure \ref{fig:schematic}.  This late-time shift influences the entire abundance distribution as Cadmium is the
most abundant element in this simulation---these dissociations therefore emit a significant number of neutrons
that are subsequently captured throughout the nuclear network.  In the top panel of Figure \ref{fig:capture} we show the
number of neutrons captured relative to the baseline simulation for both the $A=130$ peak and the region above the
$A=130$ peak.  The solid line curve decreases sharply, showing that less neutrons are captured in the $A=130$ peak as
compared to the baseline simulation, due to the photo-dissociation effect.  There is a corresponding increase in the
number of neutrons captured elsewhere.

Returning to the top panel of Fig. \ref{fig:schematic} we see that $S_{n,actual}$ and $S_{n,eq}$ cross at a slightly later
time. At this point $\lambda_\beta > \lambda_n,\lambda_\gamma$ so the increase in $S_{n,actual}$ is primarily due to beta
decay.  Nevertheless, some capture and release of neutrons is still possible, with $\lambda_n > \lambda_\gamma$ due to the
falling temperature.  Therefore, a few nuclei that are highly abundant can drain significant numbers of the remaining
neutrons from the system.  For example, at late times, the nucleus $^{131}{\rm Sn}$ becomes highly populated in the
baseline simulation because it is on the beta decay chain from a waiting point nucleus at late times, as shown in the
bottom panel of Figure \ref{fig:schematic}.  The neutron capture rate from $^{131}{\rm Sn}$ controls the number of neutrons
that are captured on the way to $^{132}{\rm Sn}$.  The impact of this drain of neutrons is shown in the bottom panel of
Figure \ref{fig:capture}.  This figure shows the number of neutrons captured in a simulation where the neutron capture rate
of $^{131}{\rm Sn}$ is increased by a factor of 100 relative to the baseline simulation.  In the $A=130$ peak more neutrons
are captured and so correspondingly less are captured in the region above the $A=130$ peak, i.e. the rare earth and $A=195$
regions. This is the same effect as described in \cite{Beu08b}.

Changing neutron capture rates creates a situation where more or less neutrons are available for capture across the
system at different times during the freeze-out process.  This causes shifts in the final abundance pattern that
depend on the relative strength and timing of the two effects.  We have shown two specific examples, but there are a
number of rates that can have these same effects.  In order to quantify the impact of individual neutron capture
rates, we calculate an average fractional change, $F$, by summing the differences of all mass fractions, $X$, of all
nuclear species
\begin{equation}
F =  100 \,  \sum_{A}  \left| X_{A,baseline} - X_A \right|
\end{equation} 
for simulations in which only one rate is varied.

\begin{figure}
\includegraphics[width=16cm]{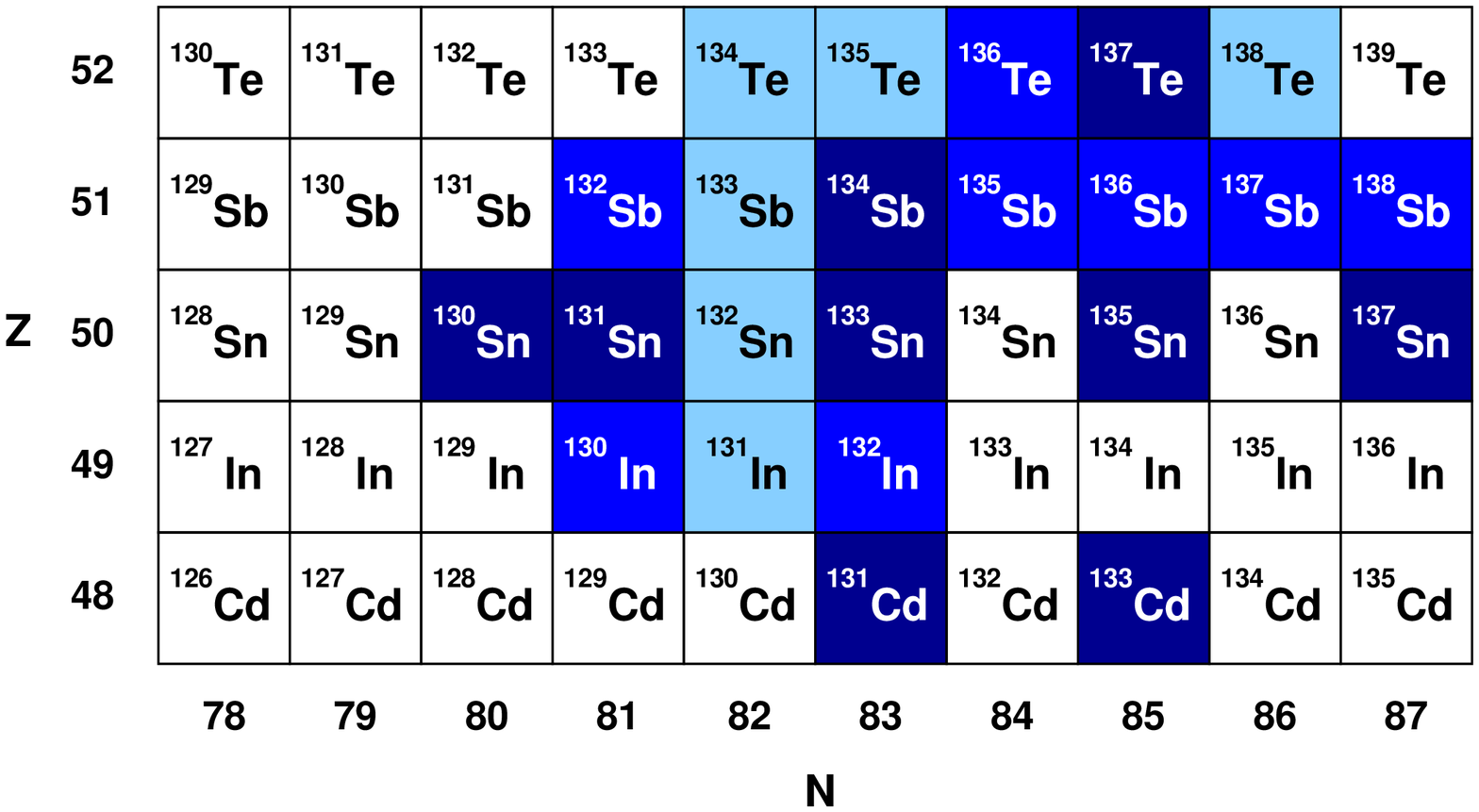}
\caption{Shows the nuclei which have neutron capture rates 
which effect a F $\approx$ 5\% or more abundance change for a rate increase of a factor of 10 (dark color
squares), a factor of 50 (medium color squares), and a factor of 100 to 1000 (light color squares).
\label{fig:colorchart}}
\end{figure}

Using different nuclear data or different astrophysical conditions can cause shifts in the path, i.e. different
nuclei are populated during freeze-out from equilibrium.  Therefore, different neutron capture rates can become
important. We study the results from a number of different simulations in order to determine the range of rates which
can play a role. We perform twelve baseline simulations using six different nuclear data sets:
\begin{itemize}
\item FRDM masses and rates, and M\"oller et al. \cite{Mol97} $\beta$-decay rates
\item FRDM masses and rates, and M\"oller et al. \cite{Mol03} $\beta$-decay rates
\item as above but with experimental data \cite{NNDC08} included 
\item ETFSI masses and rates, and M\"oller et al. \cite{Mol97} $\beta$-decay rates
\item ETFSI masses and rates, and M\"oller et al. \cite{Mol03} $\beta$-decay rates
\item as above but with experimental data \cite{NNDC08} included 
\end{itemize}
and two different outflow timescales in the wind model, $\tau = 0.1$ s and $\tau = 0.3$ s. For each of the twelve baseline
simulations we perform 250 additional simulations in which the neutron capture rate of each of the 50 nuclei in Figure
\ref{fig:colorchart} is varied by a factor of 10, 50, 100, 500, or 1000.  The darkest color squares in Figure
\ref{fig:colorchart} represent nuclei for which a factor of 10 in the neutron capture cross section yields a change in the
overall abundance pattern, $F$, of 5 to 20\% in at least one of the simulations.  Nuclei which require a factor of 50 in
order to produce at least a 5\% change in the abundance pattern have a medium shade.  For uncolored nuclei, no simulation
produced a 5\% change in the abundance pattern.  Increasing the rates of more than one nucleus at a time will compound the
effect, producing for example, the 43\% change seen in Figure \ref{fig:abund}.  Since odd-$N$ nuclei have smaller separation
energies and tend to fall out of equilibrium sooner than even-$N$ nuclei, there is a larger sensitivity to odd-$N$ capture
rates as shown in Figure \ref{fig:colorchart}.

Studies of astrophysical sites and conditions, future observations, and improved nuclear physics input are important
for determining the astrophysical site of the $r$-process. Presently the most lacking piece of input is on the side of
astrophysics theory.  Nevertheless, it is advantageous to be able to disentangle which effects arise from
astrophysical conditions and which arise from nuclear physics considerations. In addition to beta decay rates and
nuclear masses, neutron capture rates also play a role in determining final $r$-process abundance distributions.

This work was partially supported by the Department of Energy under contracts 
DE-FG05-05ER41398 (RS) and DE-FG02-02ER41216 (GCM). This work was partially supported by the
United States National Science Foundation under contract AST-0653376 (WRH).
Oak Ridge National Laboratory (WRH) is managed by UT-Battelle, LLC, for the U.S. Department
of Energy under contract DE-AC05-000R22725.


\clearpage

\clearpage

\begin{thebibliography}{}
\bibitem{Bur57}
  M.~E.~Burbidge, G.~R.~Burbidge, W.~A.~Fowler and F.~Hoyle,  Rev.\ Mod.\ Phys.\  {\bf 29}, 547 (1957).
\bibitem{Cam57}  
  A.~G.~W.~Cameron, Chalk River Rep. {\bf CRL-41} (1957).
\bibitem{Mey92}
  B.~S.~Meyer, G.~J.~Mathews, W.~M.~Howard, S.~E.~Woosley, and R.~D.~Hoffman, ApJ {\bf 399}, 656 (1992).
\bibitem{Woo94}
  S.~E.~Woosley, J.~R.~Wilson, G.~J.~Mathews, R.~D.~Hoffman, and B.~S.~Meyer ApJ {\bf 433}, 229 (1994).
\bibitem{Tak94}
  K.~Takahashi, J.~Witti, and H.-Th.~Janka, A\&A {\bf 286}, 857 (1994).
\bibitem{Beu08a}
  J.~Beun, G.~C.~McLaughlin, R.~Surman, and W.~R.~Hix, Phys.~Rev.~C {\bf 77}, 035804 (2008).
\bibitem{Fre99}
  C.~Frieburghaus, S.~Rosswog, and F.-K.~Thielemann, ApJ {\bf 525}, L121 (1999).
\bibitem{Mey89}
  B.~S.~Meyer, ApJ {\bf 343}, 254 (1989).
\bibitem{Sur08}
  R.~Surman, G.~C.~McLaughlin, M.~Ruffert, H.-Th.~Janka, and W.~R.~Hix, ApJ {\bf 679}, L117 (2008).
\bibitem{Sur06}
  R.~Surman, G.~C.~McLaughlin, and W.~R.~Hix, ApJ {\bf 643}, 1057 (2006).
\bibitem{Nin07}
  H.~Ning, Y.-Z.~Qian, and B.~S.~Meyer, ApJ {\bf 667}, L159 (2007).
\bibitem{Wan03}
  S.~Wanajo, M.~Tamamura, N.~Itoh, K.~Nomoto, Y.~Ishimaru, T.~C.~Beers, and S.~Nozawa, ApJ {\bf 593}, 968 (2003).
\bibitem{Arn07}
  M.~Arnould, S.~Goriely, and K.~Takahashi, Phys.~Rep. {\bf 450}, 97 (2007).
\bibitem{Sur01}
  R.~Surman and J.~Engel, Phys.~Rev.~C {\bf 64} 035801 (2001).
\bibitem{Rau05}
  T.~Rauscher, Nucl.~Phys.~A {\bf 758}, 655 (2005).
\bibitem{Far06}
  K.~Farouqi, K.-L.~Kratz, B.~Pfeiffer, T.~Rauscher, and F.-K.~Thielemann, in {\it Capture Gamma-Ray Spectroscopy and Related
Topics}, Vol. 819 of {\it American Institute of Physics Conference Series}, edited by A.~Woehr and A.~Aprahamian, pp. 419-422.
\bibitem{Gor98}
  S.~Goriely, Phys.~Lett.~B {\bf 436}, 10 (1998).
\bibitem{Gor97}
  S.~Goriely, A\&A {\bf 325}, 414 (1997).
\bibitem{Rau04}
  T.~Rauscher, in {\it The r-Process: The Astrophysical Origin of the Heavy Elements and Related Rare Isotope Accelerator Physics},
edited by Y.-Z.~Qian, E.~Rehm, and H.~Schatz, pp. 63.
\bibitem{Beu08b}
  J.~Beun, J.~Blackmon, W.~R.~Hix, G.~C.~McLaughlin, M.~Smith, and R.~Surman, J.~Phys.~G, submitted (2008).
\bibitem{Pea96}
  J.~M.~Pearson, R.~C.~Nayak, and S.~Goriely, Phys.~Lett.~B {\bf 387}, 455 (1996).
\bibitem{Mol95}
  P.~M{\"o}ller, J.~R.~Nix, W.~D.~Meyers, and W.~J.~Swiatecki, Atomic Data and Nuclear Data Tables {\bf 59}, 185 (1995).
\bibitem{Gor00}
  S.~Goriely (unpublished) (2000).
\bibitem{Rau00}
  T.~Rauscher and F.-K.~Thielemann, Atomic Data and Nuclear Data Tables {\bf 74}, 1 (2000).
\bibitem{Mol03}
   P.~M{\"o}ller, B.~Pfeiffer, and K.-L.~Kratz, Phys.~Rev.~C {\bf 67}, 055802 (2003).
\bibitem{Mol97}
   P.~M{\"o}ller, J.~R.~Nix, and K.-L.~Kratz, Atomic Data and Nuclear Data Tables {\bf 66}, 131 (1997).
\bibitem{NNDC08}
   N.~N.~D.~Center, http://www.nndc.bnl.gov/chart/.
\end{thebibliography}
\end{document}